\documentclass[conference,letterpaper]{IEEEtran}
\IEEEoverridecommandlockouts
\usepackage{amsmath, amssymb, dsfont, mathrsfs}
\usepackage{algorithm}
\usepackage[noend]{algpseudocode}
\usepackage{float}
\usepackage{mdframed}
\usepackage{graphicx}
\usepackage{soul}
\usepackage{enumitem}

\title{Group Testing with Runlength Constraints for Topological Molecular Storage
}

\author{
	\IEEEauthorblockN{Abhishek Agarwal, Olgica Milenkovic and Srilakshmi Pattabiraman}
	\IEEEauthorblockA{Coordinated Science Laboratory\\ University of Illinois, Urbana-Champaign\\
	\{abhiag, milenkov, sp16\}@illinois.edu }
\and
\IEEEauthorblockN{Jo\~ao Ribeiro}
\IEEEauthorblockA{Department of Computing\\ Imperial College London\\
	j.lourenco-ribeiro17@imperial.ac.uk
}
}

\date{}

\newcommand{\cS}{\mathcal{S}}
\newcommand{\cT}{\mathcal{T}}

\newcommand{\supp}{\mathsf{supp}}

\newcommand{\eps}{\epsilon}

\newcommand{\Bin}{\mathsf{Bin}}
\newcommand{\wgt}{\mathsf{wgt}}
\newcommand{\Ber}{\mathsf{Ber}}
\newcommand{\cW}{\mathcal{W}}

\newtheorem{thm}{Theorem}

\newtheorem{lem}[thm]{Lemma}

\newtheorem{defn}[thm]{Definition}

\newcommand{\M}{\mathsf{M}}

\newcommand{\bits}{\{0,1\}}

\newcommand{\Dec}{\mathsf{Dec}}

\newcommand{\pfail}{p_{\mathsf{fail}}}
\newcommand{\pcoll}{p_{\mathsf{coll}}}

\let\originalleft\left
\let\originalright\right
\renewcommand{\left}{\mathopen{}\mathclose\bgroup\originalleft}
\renewcommand{\right}{\aftergroup\egroup\originalright}

\newenvironment{proof}{\begin{IEEEproof}}{\end{IEEEproof}}

\usepackage{color}

\usepackage{color}

\usepackage{url}
\begin{document}
	
	\maketitle
	
\begin{abstract}
Motivated by applications in topological DNA-based data storage, we introduce and study a novel setting of Non-Adaptive Group Testing (NAGT) with \emph{runlength constraints} on the columns of the test matrix, in the sense that any two $1$'s must be separated by a run of at least $d$ $0$'s.
We describe and analyze a probabilistic construction of a runlength-constrained scheme in the zero-error and vanishing error settings, and show that the number of tests required by this construction is optimal up to logarithmic factors in the runlength constraint $d$ and the number of defectives $k$ in both cases.
Surprisingly, our results show that runlength-constrained NAGT is not more demanding than \emph{unconstrained} NAGT when $d=O(k)$, and that for almost all choices of $d$ and $k$ it is not more demanding than NAGT with a column Hamming weight constraint only.
Towards obtaining runlength-constrained \emph{Quantitative} NAGT (QNAGT) schemes with good parameters, we also provide lower bounds for this setting and a nearly optimal probabilistic construction of a QNAGT scheme with a column Hamming weight constraint.
\end{abstract}

\section{Introduction}\label{sec:intro}

Group testing is a pooling scheme first introduced by Dorfman~\cite{Dor43} for the purpose of testing individuals for diseases.
Since its inception, the problem and its subsequent solutions
have found a number of applications in bioinformatics (see~\cite{Dam06,emad2014semiquantitative} and references therein), information and coding theory~\cite{wolf1985born,KS64}, and many other disciplines, as outlined in~\cite{DH00,AJS19}.

In classical Non-Adaptive Group Testing (NAGT), one is concerned with the following question: Given a collection of $n$ objects of which $k \leq n$ are ``defective,'' devise a testing strategy that uses the smallest possible number of tests to identify the defectives. A test is allowed to involve an arbitrary number of objects from the pool and returns a positive answer if at least one of the objects involved is defective. 
The tests are usually summarized in what is referred to as a \emph{test matrix} -- a $t \times n$ binary matrix $M$ in which the rows correspond to the tests while the columns correspond to the test objects.
In other words, we have $M_{ij}=1$ if and only if item $i$ participates in the $j$-th test.
The set of $n$ items is usually described by a sparse vector with at most $k$ non-zero entries corresponding to the defectives.
Research in this area was kickstarted by the seminal early works of Kautz and Singleton~\cite{KS64} and D'yachkov and Rykov~\cite{DR82}.
Currently, we know explicit NAGT schemes requiring $O(k^2 \log n)$ tests in the zero-error setting~\cite{PR08} (while the probabilistic method shows the existence of such schemes with $O(k^2 \log(n/k))$ tests), and randomized schemes requiring $O(k\log n)$ tests in the average-case (i.e., vanishing error) setting with simple decoding algorithms (e.g., see~\cite{ABJ14,JAS19}).
In both cases, the number of tests is optimal up to an $O(\log k)$ factor~\cite{DR82}.

In a different context, group testing was recently shown to increase the storage density of \emph{topological} DNA-based data storage~\cite{tabatabaei2019dna}. In such a system, nanoscopic holes are punched into the sugar-phosphate backbone of one strand of a double-stranded DNA molecule. A ``hole'' indicates the value 1 while the absence of a hole indicates the value 0. Multiple copies of the same native DNA strands, referred to as registers, are punched to bear different user signatures. These are mixed together according to a group testing scheme within one pool and subsequently stored in a single microwell. The mixing process allows for using only one microwell per pool rather than using multiple microwells for individual registers thereby reducing the implementation cost.

\begin{figure}[htb]
	\centerline{\includegraphics[width=8.0cm]{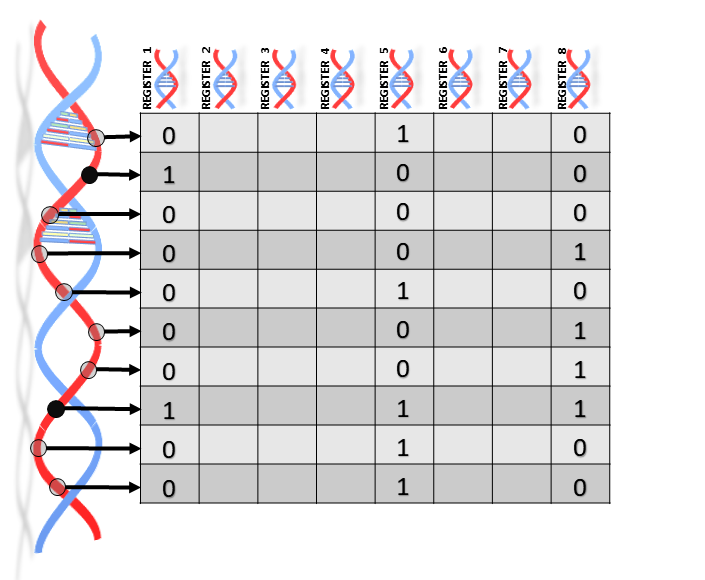}}
	\vspace{-0.1in}
	\caption{DNA Punchcards for molecular data storage. The rows index the potential nicking sites, while the columns index the DNA strands used in the mixture. Two ones in a column delineating a run of zeros correspond to a DNA fragment whose length has to be sufficiently long. Note that while Register 1 obeys a runlength constraint of $d \leq 5$, Registers 5 and 8 include two consecutive $1$s which may lead to readout errors. }
	\label{fig:punchcards}
\end{figure}

One constraint that arises in the above described group testing scheme is 
a runlength constraint for zeros between pairs of 1s on the same DNA strand, as depicted in Figure~\ref{fig:punchcards}. This constraint is associated with the quality of the readout as it is required to 
place nicks at sufficiently large distance from each other~\cite{tabatabaei2019dna}. 
We propose to address this problem by introducing a new \emph{runlength limited group testing paradigm}.

Runlength limited group testing represents a simplification of the actual mixture identification problem as the readout process also provides information about the \emph{distance} between two 1s in a register as well as the \emph{number} of DNA fragments of each type~\cite{tabatabaei2019dna}. The counts may be exploited through runlength-constrained \emph{Quantitative} NAGT (QNAGT)~\cite{SS63, Lin75}.

\subsection{Problem Setting and Basic Definitions}\label{sec:probsetting}

The Hamming weight of vector $x$ is denoted by $\wgt(x)$. A vector $x$ is said to be $k$-sparse if $\wgt(x)\leq k$.
We formulate the problem by assuming that there are $n$ items (registers), out of which $k$ are defective (i.e., included in the pool) and represented by a $k$-sparse vector $x$.
The set of non-adaptive tests to be performed is represented by a (possibly random) binary $t\times n$ test matrix $M$, with $M_{ij}=1$ if and only if item $i$ participates in the $j$-th test, and is $0$ otherwise.
In NAGT,  the test outcomes for the input $x$ are obtained through a logical OR operation, $M\odot x:=\bigvee_{j:x_j=1} M_{\cdot j}$. 
In contrast, the test outcomes in QNAGT equal $Mx$. 
In the testing scheme, we allow for all-$0$ rows in $M$ that correspond to noninformative tests but simplify our analysis.

We say that the matrix $M$ is \emph{$d$-runlength constrained} if for every $j\in [n]$ there is a $0$-run of length at least $d$ between any two $1$'s in $M_{\cdot j}$. Moreover, we say that $M$ is \emph{$w$-constrained} if $\wgt(M_{\cdot j})\leq w$ for every $j\in[n]$.
Observe that every $d$-runlength constrained matrix is also $w$-constrained for $w=\frac{t}{d+1}$.
For simplicity, throughout this work we assume that $d+1$ divides $t$. If this is not the case, we simply add up to $d$ all-$0$ tests to the scheme. As a result, our upper and lower bounds may change by at most by an additive factor of $d$.

Next, we discuss two error regimes of practical interest.

\subsubsection{The zero-error setting} We say that the test matrix $M$ represents a zero-error NAGT scheme if $M\odot x\neq M\odot x'$ for all $k$-sparse vectors $x\neq x'$.
Our goal is to design $d$-runlength zero-error NAGT schemes with $t$ as small as possible, given $n$, $k$, and $d$.
To this end, we find the following definition useful.
\begin{defn}[$k$-disjunct matrices]
A binary matrix $M$ is said to be \emph{$k$-disjunct} if the support of the bit-wise union of any collection of up to $k$ columns of $M$ does not 
contain the support of any other column of $M$.
\end{defn}

In particular, every $k$-disjunct matrix corresponds to a zero-error NAGT scheme, and every zero-error NAGT scheme is $(k-1)$-disjunct.
Moreover, $k$-disjunct matrices $M$ have an efficient deterministic decoding procedure $\Dec$ such that $\Dec(M,M\odot x)=x$.
Consequently, our goal is to design $d$-runlength $k$-disjunct matrices with small $t$.

In zero-error $d$-runlength (resp.\ $w$-constrained) QNAGT, the goal is to design a $d$-runlength (resp.\ $w$-constrained) binary $t\times n$ matrix $M$ such that $Mx\neq Mx'$ for all $k$-sparse vectors $x\neq x'$ with $t$ as small as possible.

\subsubsection{Average-case setting} Here, one only aims to ensure that the \emph{average} decoding error probability (over the randomness of the test matrix and uniform sampling of a set of $k$ defectives) vanishes with $n$.
More precisely, an average-case NAGT scheme is described by a random $t\times n$ binary matrix $\mathsf{M}$ along with a deterministic decoding procedure $\Dec$ such that
\begin{equation}\label{eq:avgcase}
	\frac{1}{\binom{n}{k}}\sum_{x:\wgt(x)=k}\mathbf{P}_{\mathsf{M}} \{ \Dec(\mathsf{M},\mathsf{M}\odot x)\neq x\}= o_n(1).
\end{equation}
Our goal is to design \emph{$d$-runlength} average-case NAGT schemes with small $t$, i.e., average-case NAGT schemes with $\mathsf{M}$ such that every valid fixing $\mathsf{M}=M$ is $d$-runlength constrained.

The definition of an average-case $d$-runlength (or $w$-constrained) QNAGT scheme is analogous to the definition above with $\Dec(\mathsf{M},\mathsf{M}\odot x)$ replaced by $\Dec(\mathsf{M},\mathsf{M} x)$ in~\eqref{eq:avgcase}.

\subsection{Related Work}\label{sec:relwork}

To the best of our knowledge this is the first line of work to consider runlength-constrained NAGT. Some recent work focused on other constrained versions of NAGT \emph{sparse} NAGT~\cite{Mac96,GGJZ19,IKO19,GRV19} which relates to our formulation. In sparse NAGT, columns and rows of the test matrix are required to satisfy certain Hamming weight constraints. Clearly, our runlength constraint on the columns of the test matrix also induces a weight constraint: Indeed, a $d$-runlength constraint on the columns of a $t\times n$ matrix induces a $w=\frac{t}{d+1}$ weight constraint on the columns as well. The weight constraints imposed by runlengths and those studied in the works mentioned above are, however, qualitatively different. In the latter, the weight constraint depends on $k$ and $n$ only, while in our case the weight constraint is a linear fraction of the number of tests (for fixed $k$ and $n$). Therefore, our results are incomparable with those of sparse NAGT.

Furthermore, starting with the work of S{\" o}derberg and Shapiro~\cite{SS63}, several works have studied the minimum number of tests required for \emph{unconstrained} QNAGT as a function of $k$ and $n$. In particular, Lindstr{\" o}m~\cite{Lin75} provided an elegant explicit, asymptotically optimal QNAGT scheme for the case $k=n$. More recently, the optimal number of tests for $k$ linear in $n$ was determined in~\cite{SC17Phase,ERKZJ19}, and for $k$ sublinear in $n$ in~\cite{GHKL19, KKHNS19}. The latter work also described efficient constructions of nearly optimal QNAGT schemes (we note that~\cite{GHKL19} allows non-binary test matrices, while all other works mentioned deal with binary test matrices only).
Our results on QNAGT can be seen as a natural extension of the problem studied above to a setting with a column runlength constraint or weight constraint.

Finally, for the motivating application, the work on semiquantitative group testing that generalizes Lindstr{\" o}m QNAGT~\cite{emad2014semiquantitative} is also of interest as it allows for handling test-dependent noise in the measurements.

\subsection{Our Contributions}\label{sec:contributions}

We briefly summarize our main results below:
\begin{itemize}
	\item \emph{Nearly-optimal runlength-constrained NAGT:} We present a probabilistic construction of a $d$-runlength NAGT scheme using $t=O((d+k)k\log(n/k))$ tests in the zero-error setting and $t=O((d+k)\log(n/k))$ tests in the average-case setting. Moreover, we derive lower bounds that show the number of tests above are optimal up to $O(\log(d \, k))$ and $O(\log(d/k))$ factors in the zero-error and average-case settings, respectively.
	
	\item \emph{Nearly optimal weight-constrained QNAGT:} As a significant step towards designing good runlength-constrained QNAGT schemes, we analyze a probabilistic construction of weight-constrained QNAGT, and derive complementary lower bounds that show that our construction is order-optimal for a large range of parameters. Note that these lower bounds also hold for runlength-constrained QNAGT.
\end{itemize}

Two interesting consequences of our results for NAGT are that (i) When $d=O(k)$, runlength-constrained NAGT is not more restrictive than \emph{unconstrained} NAGT, and (ii) For essentially \emph{all} $d$ and $k$, runlength-constrained NAGT is not more restrictive than \emph{weight-constrained} NAGT, which is a significantly weaker constraint.

\subsection{Notation}
Random variables are denoted by uppercase letters such as $X$, $Y$, and $Z$, while sets are denoted by uppercase calligraphic letters such as $\cS$ and $\cT$. The set $\{1,2,\dots,m\}$ is denoted by $[m]$.
The $i$-th row of $M$ is denoted by $M_{i\cdot}$ and its $j$-th column by $M_{\cdot j}$.
The support of a vector $x$
is denoted by $\supp(x)$; $\log$ stands for the base-2 logarithm, while $h(\cdot)$ stands for the binary entropy function. The R\'enyi entropy of order $2$ of $X$, also known as the \emph{collision entropy}, is denoted by $H_2(X)$.


\section{Probabilistic Construction of Runlength-Constrained Schemes}\label{sec:probconst}

We start our discussion with a simple NAGT construction that satisfies an arbitrary runlength constraint $d$. Let $\bar{M}_{t' \times n}$ be a given test matrix. We construct a test matrix $M_{t \times n}$ with $t = dt'$ from $\bar{M}_{t' \times n}$ by introducing $d$ all-$0$ rows between every two rows of $M$. Clearly, $M$ is a valid NAGT scheme which satisfies the given runlength constraint. 
We can instantiate this construction with the best explicit~\cite{PR08} and probabilistic~\cite{KS64} constructions of $k$-disjunct matrices. In the explicit setting, we obtain an NAGT scheme with runlength constraint $d$ using $t=\Theta(dk^2\log n)$ tests. In the probabilistic setting, we obtain an NAGT scheme with runlength constraint $d$ using $t=\Theta(dk^2\log(n/k))$ tests. In what will be clear from Theorems~\ref{thm:ze01} and~\ref{thm:ze02}, we show that our scheme based on a probabilistic construction offers significant reductions in the number of tests compared to this simplistic scheme.

One may also ask whether the standard probabilistic construction in which each entry of the test matrix is i.i.d.\ Bernoullie with some probability $p$ (which yields nearly optimal \emph{unconstrained} NAGT schemes with high probability for $p=\Theta(1/k)$) also leads to a runlength-constrained NAGT scheme. However, unless $p$ is very small (in which case the scheme's parameters are far from optimal),
the resulting test matrix will have several pairs of consecutive 1s with high probability, and hence will not satisfy any $d$-runlength constraint for $d\geq 1$.
As a result, we must consider new probabilistic constructions to obtain good parameters.

\subsection{The Zero-Error Setting}

Algorithm $\mathsf{RandMatrix}(t, n, d, k, \alpha)$ describes our scheme to construct a matrix $M_{t \times n }$, that with high probability, is $k$-disjunct with a weight bound $\alpha$.
\vspace{3pt}
\hrule \vspace{0.5pt} \hrule
\vspace{3pt}
$\mathsf{RandMatrix}(t, n, d, k, \alpha)$
\vspace{3pt}
\hrule \vspace{0.5pt} \hrule
\vspace{3pt}
$\mathsf{Input:}$ Runlength constraint $d$, weight constraint $\alpha \leq t/(2d+1)$.

$\mathsf{Output:}$ $k$-disjunct matrix $M_{t \times n}$
\vspace{3pt}
\hrule 
\vspace{3pt}
Each column $M_{.j}$, $ 1\leq j \leq n$ is constructed identically and independently via the following procedure:
\begin{enumerate}[leftmargin=3\parindent]
	\item Set the list $I \gets (1,2,\ldots,t)$. Set $w \gets 1$.
	\item Pick an index $i$ uniformly at random from the list $I$. 
	\item $M_{ij} \gets 1$; $w \gets w+1$. 
	\item Let $U$ be the set of $2d$ indices symmetrically and cyclically ``surrounding'' $i$ in $I$. 
	
	For example, if $d=2$ and in the current iteration 
	
	$I = (1,2,3,4,5,6,7,13,14)$,  $j = 13$, then 
	
	$U \gets (6,7,14,1)$ as 6 and 7 precede 13, while 14 and 1 succeed 13 in cyclic order.
	\item For all $i' \in U$, $M_{i'j} \gets 0$. 
	\item Update $I \gets I-U$.
	\item Iterate starting from step 2) as long as $w \leq \alpha$.
	\item For all $i' \in I$, $M_{i'j} \gets 0$.
\end{enumerate}
\vspace{3pt}
\hrule \vspace{0.5pt} \hrule
\vspace{3pt}

\begin{thm}\label{thm:ze01}
	Let $\alpha = k \log (n/k)$ and $t = \Theta (dk \log(n/k)+ k^2 \log(n/k))$. $\mathsf{RandMatrix}(t, n, d, k, \alpha)$ returns a $k$-disjunct matrix $M$ that satisfies a $d$ runlength constraint with probability at least $1-O(1/n)$. 
\end{thm}
\begin{proof}
	Given a matrix $M_{t \times n}$ that is an output of the $\mathsf{RandMatrix}(t, n, d, k, \alpha)$ with the parameters as set in Theorem~\ref{thm:ze01}, we show that the the probability of $M_{t \times n}$ not being $k$-disjunct is at most $O(1/n)$. 
	Let $v$ denote an arbitrary column of $M$ and let $\supp(v)=\{{i_1, \dots, i_\alpha\}}$, where the weight of $v$ is $\alpha \leq t/(2d +1)$. Let $V=\{{v_1, \dots, v_k\}}$ denote $k$ columns of $M$ that differ from $v$. To avoid notational clutter, we write $\supp(V)$ instead of $\cup_{i=1}^k \supp(v_i)$. We wish to show that $\supp(v) \not \subseteq \supp(V)$ with high probability over the randomness of the algorithm.

	The probability that an index $i_{\ell}$ is covered in the support of $V$ satisfies
	\begin{multline}
	\mathbf{P}\{i_\ell \in \supp(V)\}  \leq \sum_{j =1}^{k} \mathbf{P}\{i_\ell \in \supp(v_j)\} \\= k \cdot \mathbf{P}\{i_\ell \in \supp(v_1)\},  
	\end{multline} which is an immediate consequence of the i.i.d.\ assumption on the columns $v_i$'s and the union bound. Also, by the chain rule of probability we have
	\begin{align}
	&\mathbf{P}\{i_1, \dots, i_\alpha \in \supp(V)\}= \nonumber \\
	&\mathbf{P}\{i_1 \in \supp(V) \} \, \mathbf{P}\{i_2 \in \supp(V) | i_1 \in \supp(V) \}  \nonumber \\
	& \quad \cdots  \mathbf{P}\{i_\alpha \in \supp(V)|i_1, \dots, i_{\alpha -1}\in \supp(V)\}. \label{eq:ze02} 
	\end{align}
	We start by deriving an upper bound for the first term in~\eqref{eq:ze02}. Note that the probability of the event $\{{i_1 \in  \supp(v_1)\}}$ is the probability that the following events occur: Index $i_1$ is picked at the first step, \emph{or} an index outside the symmetric $d$-neighborhood of the index $i_1$ is picked at the first step and $i_1$ is picked at the first step, \emph{or} indices outside the $d$- neighborhood of index $i_1$ are selected at the first two steps and $i_1$ is picked at the third step, etc. Thus,
	\begin{align}
	&\mathbf{P}\{i_1 \in \supp(v_1)\} = \frac{1}{t} + \frac{t-(2d+1)}{t} \cdot \frac{1}{t - (2d+1)} + \dots + \nonumber \\
	&\frac{t-(2d+1)}{t} \cdots \frac{t-(2d+1)(\alpha-2)}{t-(2d+1)(\alpha-1)} \cdot \frac{1}{t-(2d+1)(\alpha-2)} \nonumber \\ 
	&= \frac{\alpha}{t}. \nonumber
	\end{align}
	Therefore, 
	\begin{align}
	\mathbf{P}\{i_1 \in \supp(V)\} \leq k \cdot \frac{\alpha}{t}.
	\end{align}
	To find an upper bound for the second term in~\eqref{eq:ze02} we proceed as follows. 
	\begin{align}
	&\mathbf{P}\{i_2 \in \supp(V) | i_1 \in \supp(V) \} \label{eq:tower} \\
	&= \mathbf{P}\{i_2 \in \cup_{j =1}^k \supp(v_j) | i_1 \in \supp(V) \} \nonumber\\
	& \leq  \sum_{j =1}^k \mathbf{P}\{i_2 \in \supp(v_j) | i_1 \in \supp(V) \}\nonumber \\
	&= k\cdot \mathbf{P}\{i_2 \in \supp(v_1) | i_1 \in \supp(V) \} \nonumber \\
	&= k\, ( q\cdot \mathbf{P}\{i_2 \in \supp(v_1) | i_1 \in \supp(V), i_1 \in \supp(v_1) \} \nonumber\\ 
	& + (1-q)\cdot \mathbf{P}\{i_2 \in \supp(v_1) | i_1 \in \supp(V), i_1 \not \in \supp(v_1) \}), \nonumber 
	\end{align}
	where $q = \mathbf{P}\{i_1 \in \supp(v_1) |i_1 \in \supp(V)\}$. 
	Furthermore, since indices $i_1, i_2 \in \supp(v)$ are at least $d$ apart, 
	\begin{align}
	&\mathbf{P}\{i_2 \in \supp(v_1) | i_1 \in \supp(V), i_1 \not \in \supp(v_1) \}  \nonumber\\
	& \leq \frac{1}{t-1} + \frac{t-1 -(2d+1)}{t-1} \cdot \frac{1}{t-1 -(2d+1)} + \cdots \nonumber \\    
	& \quad + \frac{t-1 -(2d+1)}{t-1}  \cdots \frac{t-1 -(\alpha-1)(2d+1)}{t-1 - (\alpha -2)(2d+1)} \cdot \nonumber\\
	& \quad \frac{1}{t-1 -(\alpha-1)(2d+1)} = \frac{\alpha}{t-1}. \nonumber
	\end{align}
	Let $E_{\{i_1,\ell\}}$ denote the event that index $i_1$ was the $\ell^{\text{th}}$ index to be set to $1$ in $v$. Then,
	\begin{align}
	&\mathbf{P}\{i_2 \in \supp(v_1) | i_1 \in \supp(V), i_1 \in \supp(v_1) \} \nonumber\\
	&= \sum_{\ell=1}^{\alpha} \mathbf{P}\{E_{\{i_1,\ell\}}\} \, \mathbf{P}\{i_2 \in \supp(v_1) |  i_1 \in \supp(v_1), E_{\{i_1,\ell\}} \}\nonumber\\
	&= \sum_{\ell=1}^{\alpha} \mathbf{P}(E_{\{i_1,\ell\}}) \cdot g(\ell), \nonumber
	\end{align}
	where $g(1)$ equals
	\begin{align}
	&0 + \frac{1}{t-(2d+1)} + \frac{t-2(2d+1)}{t-(2d+1)}\cdot \frac{1}{t-2(2d+1)} + \cdots \nonumber \\
	& + \frac{t-2(2d+1)}{t-(2d+1)} \cdots \frac{t-(\alpha-2)(2d+1)}{t-(\alpha -3)(2d+1)} \cdot \nonumber \\ 
	& \frac{1}{t-(\alpha-2)(2d+1)} = \frac{\alpha-1}{t-(2d+1)}. \nonumber
	\end{align}
	Analogously, $g(2) = \dots = g(\alpha) = \frac{\alpha-1}{t-(2d+1)}.$
	Hence, 
	\begin{align}
	\mathbf{P}&\{i_2 \in \supp(V) | i_1 \in \supp(V) \} \label{eq:together} \\ \nonumber
	&=  k\, ( q \cdot \mathbf{P}\{i_2 \in \supp(v_1) | i_1 \in \supp(V), i_1 \in \supp(v_1) \} \\ \nonumber 
	& + (1-q) \cdot \mathbf{P}\{i_2 \in \supp(v_1) | i_1 \in \supp(V), i_1 \not \in \supp(v_1) \})\\ \nonumber
	&\leq k \cdot \max \left( \frac{\alpha-1}{t-(2d+1)}, \frac{\alpha}{t-1} \right) < k \cdot \frac{\alpha}{t-(2d+1)}. \nonumber
	\end{align}
	Through analysis analogous to equations \eqref{eq:tower} and \eqref{eq:together}, we conclude that 
	\begin{align}
	\mathbf{P}&\{i_\ell \in \supp(V) | i_1, \dots, i_{\ell -1} \in \supp(V)\}\nonumber \\
	&< k\cdot \frac{\alpha}{t- (\ell-1)(2d+1)}.
	\end{align}
		In summary, we have
	\begin{align}
	\mathbf{P}&\{i_1, \dots, i_\alpha \in \supp(V)\} \nonumber \\
	&= \mathbf{P}\{i_1 \in \supp(V)\} \mathbf{P}\{i_2 \in \supp(V) | i_1 \in \supp(V)\} \nonumber \\
	& \qquad \cdots \mathbf{P}\{i_\alpha \in \supp(V)|\ i_1, \dots, i_{\alpha -1} \in \supp(V)\} \nonumber \\
	& < \frac{k\alpha}{t} \cdot \frac{k \alpha}{t-(2d+1)}\cdots \frac{k \alpha}{t-(2d+1)(\alpha-1)} \nonumber \\
	&\leq  \left( \frac{k \alpha}{t-(2d+1)(\alpha-1)} \right)^\alpha. \nonumber 
	\end{align}
	As a result, the probability that a fixed column of the test matrix $M$ is contained in the bitwise union of any other $k$ columns of $M$ is upper bounded by
	\begin{equation*}
	\binom{n}{k} \cdot \left(\frac{k\alpha}{t-(2d+1)(\alpha-1)}\right)^\alpha.
	\end{equation*}
	Using the union bound over all $n$ columns of $M$ we see that the probability that $M$ is not $k$-disjunct is upper bounded by
	\begin{align}
	& n\cdot \binom{n}{k} \cdot \left(\frac{k\alpha}{t-(2d+1)(\alpha-1)}\right)^\alpha\nonumber\\
	&< n\cdot \left(\frac{en}{k}\right)^k \cdot \left(\frac{k\alpha}{t-(2d+1)(\alpha-1)}\right)^\alpha.\label{eq:pfailub}
	\end{align}
	Choosing the parameters as in the statement of Theorem~\ref{thm:ze01} proves that this probability is bounded by $O(1/n)$. 
\end{proof}

In Section \ref{sec:lb}, we show that this construction is optimal up to logarithmic factors in $d$ and $k$.

\subsection{Average-Case Setting}

We now analyze the performance of our probabilistic construction with respect to \emph{average-case} NAGT. 
We will sample the test matrix via the $\mathsf{RandMatrix}(t, n, d, k, \alpha)$ procedure, and will focus on $\mathsf{COMP}$ decoding~\cite{CCJS11,Ald17}. 
In particular, we can achieve an error probability $O(1/n)$. We shall see in Section~\ref{sec:lb} that the number of tests required by $\mathsf{RandMatrix}(t, n, d, k, \alpha)$ under
$\mathsf{COMP}$ decoding is order-optimal for all $k$ and $d$.

\begin{thm}\label{thm:ze02}
	Let $\alpha = k \log (n/k)$, $t = \Theta(d\log(n/k)+k\log(n/k))$ and $c<1$ be some positive constant. If the $k=O(n^c)$ defectives are chosen uniformly at random, then $\mathsf{RandMatrix}(t, n, d, k, \alpha)$ returns a matrix $M$ that decodes the defectives correctly with probability at least $1- O(1/n)$. 
\end{thm}

\begin{proof}
	Following a similar analysis as that used to prove Theorem \ref{thm:ze01}, we find that
	\begin{align*}
	\mathbf{P}_{\mathsf{M}} \{ \Dec(\mathsf{M},\mathsf{M}\odot x)\neq x \} \leq n\cdot\left(\frac{k\alpha}{t-(2d+1)(\alpha-1)}\right)^\alpha,
	\end{align*}
	where $\Dec$ is $\mathsf{COMP}$ decoding. 
	
	Setting $\alpha = k\log (n/k)$ and $t = \Theta(d\log(n/k)+k\log(n/k))$, we obtain the desired upper bound on the failure probability of $O(1/n)$ from~\eqref{eq:avgcase}.  
\end{proof}


\section{Lower Bounds for Runlength-Constrained Schemes}\label{sec:lb}

We now derive lower bounds on the number of tests required for the runlength-constrained zero-error and average-case NAGT schemes.
Our lower bounds show that the number of tests of the probabilistic construction presented in Section~\ref{sec:probconst} is tight up to a logarithmic factor in both regimes.
The lower bounds we prove below only use the fact that a $d$-runlength constraint induces a Hamming weight constraint $w=\frac{t}{d+1}$ on the columns of the test matrix.
Therefore, they hold for all $w$-constrained NAGT schemes.
Surprisingly, this shows that runlength-constrained NAGT schemes are not worse than weight-constrained NAGT schemes.

\subsection{Zero-Error Setting}

We begin by noting that, since every zero-error NAGT test matrix for $k$ defectives is also $(k-1)$-disjunct, in the zero-error setting it suffices to prove lower bounds on the number of rows of runlength-constrained disjunct matrices.

Before we proceed with the proof of the main lower bound, we need the following definition and lemma.
\begin{defn}[Private set]
	Given a $t\times n$ matrix $M$ and $j\in[n]$, a set $\cS\subseteq [t]$ is said to be \emph{$(M,j)$-private} if $M_{ij}=1$ for all $i\in\cS$ and for every $j'\neq j$ there is an $i'\in\cS$ such that $M_{i'j'}=0$.
\end{defn}

\begin{lem}\label{lem:lb01}
	Every $d$-runlength $k$-disjunct $t\times n$ matrix must satisfy
	\begin{equation*}
	t\geq \min(n,1+k(d+1)).
	\end{equation*}
\end{lem}
\begin{proof}
	Fix a $d$-runlength $k$-disjunct $t\times n$ matrix $M$, and let $w=\frac{t}{d+1}$.
	If $k<w$, then we immediately conclude that $t\geq 1+k(d+1)$.
	To complete the proof, we show that $k\geq w$ implies that $t=n$.
	Indeed, if $k\geq w$, then every column of $M$ has weight at most $k$.
	In turn, since $M$ is $k$-disjunct, this means that for every $j\in[n]$ there is an $(M,j)$-private set $\cS_j$ of size $1$.
	Since $\cS_j\neq \cS_{j'}$ for all $j\neq j'$, it follows that $t\geq n$.
\end{proof}

We are now ready to prove the main lower bound.
In order to do this, we modify the technique based on private sets used to prove the well-known $t=\Omega\left(k^2 \frac{\log n}{\log k}\right)$ lower bound for unconstrained NAGT schemes~\cite[Section 19]{GRS18}.
\begin{thm}\label{thm:lb02}
	Suppose that $k(d+1)\geq 4$.
	Then, every $d$-runlength $k$-disjunct $t\times n$ matrix must satisfy
	\begin{equation*}
	t\geq \min\left(n,\Omega\left(\frac{k\cdot d\log n}{\log(k\cdot d)}\right)\right).
	\end{equation*}
\end{thm}
\begin{proof}
	Fix a $d$-runlength $k$-disjunct $t\times n$ matrix $M$, and let $w=\frac{t}{d+1}$.
	We may assume that $k<w$ (otherwise $t=n$ by Lemma~\ref{lem:lb01}).
	In particular, this means that $w/k>1$.
	
	We begin by showing that every $j\in[n]$ has an $(M,j)$-private set $\cS_j$ of size $|\cS_j|\leq 2w/k$.
	To establish a contradiction, suppose that there is a $j$ that does not satisfy this condition.
	Partition $\supp(M_{\cdot j})$ into $a\leq k$ subsets $\cW_1,\dots,\cW_a$ each of size at most $2w/k$.
	This is possible because $\wgt(M_{\cdot j})\leq w$.
	By assumption, no $\cW_b$ is $(M,j)$-private for $b=1,\dots,a$.
	This means that $M_{ij}=1$ for every $i\in\cW_b$ (since $\cW_b\subseteq\supp(M_{\cdot j})$), but there is a $j_b\neq j$ such that $M_{ij_b}=1$ for all $i\in\cW_b$.
	Since $\cW_1,\dots,\cW_a$ partition $\supp(M_{\cdot j})$, it follows that $M_{\cdot j_1},\dots,M_{\cdot j_a}$ cover $M_{\cdot j}$.
	This contradicts the fact that $M$ is $k$-disjunct, because $a\leq k$.
	
	Since every $j\in [n]$ has an $(M,j)$-private set of size at most $2w/k$, and each such set is private for at most one index $j$, it follows that
	\begin{equation*}
	n\leq |\{\cS\subseteq [t]: |\cS|\leq 2w/k\}|=\sum_{i=0}^{2w/k}\binom{t}{i}.
	\end{equation*}
	The standard entropy upper bound on the volume of the Hamming ball then implies that
	\begin{align}
	\log n&\leq t\cdot h\left(\frac{2w}{k\cdot t}\right)\nonumber\\
	&=t\cdot h\left(\frac{2}{k(d+1)}\right)\nonumber\\
	&\leq \frac{t}{k(d+1)}\cdot(\log\left(k(d+1)\right)-1),\label{eq:mainineq0error}
	\end{align}
	where the last inequality follows from the fact that $h(p)\leq -2p\log p$ for $p\leq 1/2$ and the assumption that $k(d+1)\geq 2$.
	Rearranging~\eqref{eq:mainineq0error} leads to the desired lower bound on $t$.
\end{proof}

Combining the lower bound from Theorem~\ref{thm:lb02} with the $t=\Omega\left(k^2\cdot \frac{\log n}{\log k}\right)$ lower bound for general NAGT schemes allows us to conclude that the probabilistic construction from Section~\ref{sec:probconst} is optimal for all regimes of $k$ and $d$ up to an $O(\log(k\cdot d))$ factor.

\subsection{Average-Case Setting}

In order to show our probabilistic construction is also nearly optimal (up to an $O(\log(d/k))$ factor) in the average-case setting, we employ a simple information-theoretic argument that stems directly from the fact that the vector of test outcomes must have relatively small Hamming weight (and hence it has low entropy).
\begin{thm}\label{thm:avglb}
	Suppose that $d\geq 2k$.
	Then, every average-case $d$-runlength NAGT scheme must have
	\begin{equation*}
		t=\Omega\left(\frac{d\log(n/k)}{\log(d/k)}\right).
	\end{equation*}
\end{thm}
\begin{proof}
	Suppose $\M$ is an average-case $d$-runlength NAGT scheme.
	Let $X$ be uniformly distributed over the set of $k$-sparse vectors, and denote the test outcomes of $\M$ on input $X$ by $Y$.

	Using the fact that $\M$ is an average-case NAGT scheme coupled with Fano's inequality, it follows that, for some $\eps=o(1)$,
	\begin{align}\label{eq:lby}
		H(Y|\M)&\geq H(X)-h(\eps)-\eps\log V(n,k)\nonumber\\
		&\geq (1-o(1)) k\log(n/k),
	\end{align}
	where $V(n,k)$ denotes the volume of the $n$-dimensional radius-$k$ Hamming ball.
	It remains to upper bound $H(Y|\M)$ appropriately.
	Fix $\M=M$.
	Then, by the runlength constraint, we have that every column of $Y$ has weight at most $\frac{t}{d+1}$.
	Therefore, since $X$ is $k$-sparse, it follows that $\wgt(Y)\leq \frac{k\cdot t}{d+1}$.
	As a result, we conclude that
	\begin{equation}\label{eq:uby}
		H(Y|\M=M)\leq \log V\left(t,\frac{k\cdot t}{d+1}\right)\leq t\cdot h\left(\frac{k}{d+1}\right)
	\end{equation}
	for every possible fixing $\M=M$.
	Combining~\eqref{eq:lby},~\eqref{eq:uby}, and the inequality $h(p)\leq -2p\log p$ valid for all $p\leq 1/2$ leads to the desired lower bound on $t$.
\end{proof}

\section{Constrained Quantitative Group Testing}\label{sec:QNAGT}

Recall that in QNAGT our goal is to design a (potentially random) matrix $M$ such that $Mx\neq Mx'$ for all pairs of $k$-sparse vectors $x\neq x'$ using as few rows as possible.
In this section, we study lower bounds for, and constructions of, such QNAGT schemes with runlength or column weight constraints.

\subsection{Lower Bounds}\label{sec:QNAGTlb}

In this section, we present our lower bounds on the number of tests of runlength-constrained QNAGT schemes.
Similarly to Section~\ref{sec:lb}, the lower bounds we prove below also hold for all QNAGT schemes with a column weight constraint.

The first lower bound follows from a non-trivial modification of the information-theoretic argument used by Lindstr\"om~\cite{Lin75} to derive lower bounds for general QNAGT schemes.
Before we proceed, we require the following lemmas.
\begin{lem}[\cite{Mad08}]\label{lem:submod}
	If $Z_1, Z_2, Z_3$ are independent $\mathbb{R}^k$-valued random variables, it holds that
	\begin{equation*}
	H(Z_1+Z_2+Z_3)-H(Z_2+Z_3)\leq H(Z_1+Z_2)-H(Z_2).
	\end{equation*}
\end{lem}

\begin{lem}[\cite{Mas88}]\label{lem:intentbound}
	Let $Z$ be an integer-valued random variable with variance $\sigma^2$. Then, we have
	\begin{equation*}
		H(X)\leq \frac{1}{2}\log(2\pi e(\sigma^2 +1/12)).
	\end{equation*}
\end{lem}

\begin{thm}\label{thm:QNAGTlb1}
	For every constant $\delta>0$, $n$ large enough, and $k\geq c_\delta\log n$ for a large enough constant $c_\delta>0$, every $d$-runlength QNAGT scheme must have
	\begin{equation*}
		t\geq \frac{2(1-\delta)k\log(1+n/k)}{\log\left(\frac{2\pi e k}{d+1}+2\right)}.
	\end{equation*}
\end{thm}
\begin{proof}
	Suppose $M$ is a $d$-runlength QNAGT scheme and fix a constant $\delta>0$.
	Consider $X\in\bits^n$ obtained by sampling each $X_i$ independently according to $\Ber\left(p=(1-\delta/2)k/n\right)$, and denote the $t$-tuple of test outcomes on input $X$ by $Y$.
	Observe that each $Y_i$ is distributed according to $\Bin(\ell_i,p)$, where $\ell_i=|\wgt(M_{i\cdot})|$.
	Moreover, the $d$-runlength constraint enforces that 
	\begin{equation}\label{eq:constelli}
	\sum_{i=1}^{t}\ell_i\leq w\cdot n,
	\end{equation}
	where $w=\frac{t}{d+1}$.

	A standard application of the Chernoff bound shows that
	\begin{equation*}
		p_{\mathsf{bad}}=\mathbf{P}\{\wgt(X)>k\} \leq 2^{-\frac{\delta^2 k}{12}}= 1/n
	\end{equation*}
	provided $k\geq \frac{12\log n}{\delta^2}$.
	Let $B$ denote the indicator random variable of the event $\wgt(X)> k$.
	Then, for large enough $n$ (and hence $k$), we have
	\begin{align}
		\sum_{i=1}^t H(Y_i)&\geq H(Y)\nonumber\\
		&\geq (1-p_{\mathsf{bad}})H(Y|B=0)\nonumber\\
		&\geq (1-p_{\mathsf{bad}})H(X|B=0)\nonumber\\
		&\geq H(X)-h(p_{\mathsf{bad}})-p_{\mathsf{bad}}\cdot n\nonumber\\
		&\geq (1-\delta)k\log(1+n/k),\label{eq:ylb}
	\end{align}
	where the last inequality holds for large enough $n$ and $k$ because $H(X)=n\cdot h(p)\geq (1-\delta/2)k\log(1+n/k)$, $h(p_{\mathsf{bad}})=o(1)$, and $p_{\mathsf{bad}}\cdot n\leq 1$.
	We proceed to show that~\eqref{eq:constelli} enforces the inequality
	\begin{equation}\label{eq:ineqd}
	\sum_{i=1}^{t} H(Y_i)\leq \frac{t}{2}\cdot\log\left(\frac{2\pi e k}{d+1}+2\right)
	\end{equation}
	Coupled with~\eqref{eq:ylb}, this immediately implies that
	\begin{equation*}
	(1-\delta)k\log(1+n/k)\leq \frac{t}{2}\cdot\log\left(\frac{2\pi e k}{d+1}+2\right),
	\end{equation*}
	which yields the desired lower bound on $t$.
	
	It remains to show that~\eqref{eq:ineqd} holds.
	Note that the inequality holds when $\ell_i=\frac{w\cdot n}{t}=\frac{n}{d+1}$ for all $i$, since for every constant $\eps>0$ we have
	\begin{equation*}
	H\left(\Bin\left(\frac{n}{d+1},p\right)\right)\leq \frac{1}{2}\cdot\log\left(\frac{2\pi e k}{d+1}+2\right)
	\end{equation*}
	for $n$ large enough by Lemma~\ref{lem:intentbound}.
	We claim that this is the maximizing assignment of the $\ell_i$'s. 
	This follows easily from the fact that
	\begin{multline}\label{eq:submod}
	H(\Bin(\ell_2,1/2))-H(\Bin(\ell_2-1,1/2))\\
	\leq H(\Bin(\ell_1,1/2))-H(\Bin(\ell_1-1,1/2))
	\end{multline}
	for all $\ell_1\leq\ell_2$, which in turn corresponds to Lemma~\ref{lem:submod} with $Z_1=W_1$, $Z_2=\sum_{i=2}^{\ell_1} W_i$, and $Z_3=\sum_{i=\ell_1+1}^{\ell_2} W_i$, where the $W_i$ are i.i.d.\ according to $\Ber(p)$.
	To prove the claim above using~\eqref{eq:submod}, fix some arbitrary assignment of the $\ell_i$'s satisfying~\eqref{eq:constelli}.
	Without loss of generality, we may assume that~\eqref{eq:constelli} is satisfied with equality.
	Then, either $\ell_i=\frac{n}{d+1}$ for all $i$, or there are $a\neq a'$ such that $\ell_a<\frac{n}{d+1}$ and $\ell_{a'}>\frac{n}{d+1}$.
	In the latter case, consider the alternative assignment $(\ell'_i)$ (satisfying~\eqref{eq:constelli} with equality) such that $\ell'_a=\ell_a+1$, $\ell'_{a'}=\ell_{a'}-1$, and $\ell'_i=\ell_i$ for $i\neq a, a'$.
	If $Y_i\sim \Bin(\ell_i,p)$ and $Y'_i\sim\Bin(\ell'_i,p)$, then~\eqref{eq:submod} immediately implies that
	\begin{equation*}
	\sum_{i=1}^{t}H(Y_i)\leq \sum_{i=1}^{t}H(Y'_i).
	\end{equation*}
	Repeating this argument until $\ell_i=\frac{n}{d+1}$ for all $i$ leads to the claim.
\end{proof}

Using the fact that the vector of test outcomes $y$ from a $d$-runlength QNAGT scheme satisfies $\wgt(y)\leq\frac{t\cdot k}{d+1}$ and $0\leq y_i\leq k$ for every $i=1,\dots,t$, a reasoning similar to that used to prove Theorem~\ref{thm:avglb} leads to the following result.
\begin{thm}\label{thm:QNAGTlb2}
	Suppose that $d\geq 2k$.
	Then, every $d$-runlength QNAGT scheme must have
	\begin{equation*}
	t=\Omega\left(\frac{d\log(n/k)}{\log(d^2/k)}\right).
	\end{equation*}
\end{thm}

The lower bounds from Theorems~\ref{thm:QNAGTlb1} and~\ref{thm:QNAGTlb2} complement each other, and can also be seen to hold (with slightly smaller leading constants) in the average-case setting.
Indeed, the lower bound from Theorem~\ref{thm:QNAGTlb2} improves upon the one from Theorem~\ref{thm:QNAGTlb1} whenever the runlength constraint $d$ is significantly larger than $2k$. 

\subsection{Towards a Tight Probabilistic Construction}

The lower bound from Theorem~\ref{thm:QNAGTlb1} holds even if we only require that the QNAGT scheme satisfy a column-weight constraint $w=\frac{t}{d+1}$.
Phrased in terms of $w$, every average-case $w$-constrained QNAGT scheme must have
\begin{equation}\label{eq:lbweight}
t\geq\frac{2(1-\delta)k\log(1+n/k)}{\log\left(\frac{kw}{t}+2\right)}.
\end{equation}
Below, we construct an average-case $w$-constrained QNAGT scheme consisting of
\begin{equation*}
	t=\frac{c\cdot  k\log(n/k)}{\log w}
\end{equation*}
tests, with $c\approx 4$ for a large range of $k$ and $w$.
According to~\eqref{eq:lbweight}, this is optimal up to a constant factor.
Before we present our construction, we present two lemmas.
\begin{lem}\label{lem:ubcoll}
	Let $X$ and $Y$ be i.i.d.\ according to $\Bin(\ell,p)$.
	If $\ell$ and $p=p(\ell)$ are such that $\ell p,\ell(1-p)\to\infty$ when $\ell\to\infty$, then
	\begin{equation*}
		\mathbf{P}\{X=Y\} \leq \frac{1}{\sqrt{2\ell p(1-p)}}
	\end{equation*}
	for $\ell$ large enough.
\end{lem}
\begin{proof}
	First, note that
	\begin{equation*}
		\mathbf{P}\{X=Y\}\leq \max_i \mathbf{P}\{ \Bin(\ell,p)=i \}.
	\end{equation*}
	Hence, it suffices to show the desired inequality for $\max_i \mathbf{P}\{ \Bin(\ell,p)=i \} $, which is achieved at $i=\lfloor (\ell+1)p\rfloor$ or $i=\lceil (\ell+1)p\rceil -1$.
	This follows from a direct application of Stirling's approximation of the factorial to estimate the relevant probabilities, which holds whenever the conditions in the lemma statement are satisfied.
\end{proof}
%

\begin{lem}\label{lem:renyidec}
	Let $X_i$ be i.i.d.\ according to some integer-valued distribution $X$ for $i=1,2,\dots$.
	Then, it holds that
	\begin{equation*}
		H_2\left(\sum_{i=1}^n X_i\right)\leq H_2\left(\sum_{i=1}^{n+1} X_i\right)
	\end{equation*}
	for every $n\geq 1$.
\end{lem}
\begin{proof}
	Note that the collision probability of a given distribution is the squared $2$-norm of its probability mass function (pmf) seen as a real-valued sequence, and that the pmf of $\sum_{i=1}^{n+1} X_i$ is the discrete convolution of the pmf's of $\sum_{i=1}^{n} X_i$ and $X$.
	The desired inequality then follows directly by applying Young's inequality for convolution. 
\end{proof}

We are now ready to describe and analyze our candidate construction of a $w$-constrained QNAGT scheme.
\begin{thm}\label{thm:wconsQNAGT}
	Given arbitrary constants $\delta\in (0,1/2)$ and $\gamma>0$, for large enough $n$, $k\leq n/e$, and $w\geq k^{1/2+\delta}$ such that $w\log w=o(k)$ and $w=e^{\omega(\log\log n)}$, there exists an average-case $w$-constrained $(t,k)$-QNAGT scheme with
	\begin{equation*}
	t=\frac{(2+1/\delta+\gamma)k(1+\log(n/k))}{\log w}.
	\end{equation*}
\end{thm}
\begin{proof}
	Consider the following process for sampling a binary $t\times n$ matrix $M$: Each entry $M_{ij}$ is i.i.d.\ according to $\Ber(p)$, with $p=\frac{w}{2t}$.
	A straightforward application of the Chernoff bound coupled with the choice of $w$ shows that $\wgt(M_{\cdot j})> w$ holds with probability at most $o(1/n)$ for every $j$.
	Therefore, a union bound over all $n$ columns implies that $M$ satisfies the column weight constraint with probability $1-o(1)$.
	As a result, in order to show the existence of an average-case $w$-constrained $(t,k)$-QNAGT scheme, it now suffices to prove that for every vector $x\in\bits^n$ of weight $k$ we have
	\begin{equation}\label{eq:avgcasecond}
		p_{\mathsf{fail}}(x):=\mathbf{P}\{ \exists x'\neq x: Mx=Mx'\wedge \wgt(x')=k\}=o(1),
	\end{equation}
	where the probability is taken over the randomness of $M$.
	
	We proceed to show~\eqref{eq:avgcasecond}.
	Fix an arbitrary $x\in\bits^n$ of weight $k$,
	and let $\pcoll(\ell,p)$ denote the collision probability of a $\Bin(\ell,p)$ distribution.
	Then, a union bound over all $x'\neq x$ of weight $k$ yields
	\begin{equation}\label{eq:unionboundpfail}
		\pfail(x)\leq \sum_{\ell=1}^k\binom{k}{\ell} \binom{n-k}{\ell} \pcoll(\ell,p)^t.
	\end{equation}
	The inequality in~\eqref{eq:unionboundpfail} is obtained by noting that $M(x-x')=0$ if and only if
	\begin{equation*}
		\sum_{j: x_j=1\wedge x'_j=0} M_{\cdot j}=\sum_{j': x_j=0\wedge x'_j=1} M_{\cdot j'}.
	\end{equation*}
	Moreover, the two sums above are i.i.d., and $\sum_{j=1}^\ell M_{ij}$ are i.i.d.\ according to $\Bin(\ell,p)$ for every $i\in[t]$.
	We now divide the right-hand side of~\eqref{eq:unionboundpfail} into two parts which are upper bounded by $o(1)$ in different ways.
	First, note that for large enough $n$ we have
	\begin{align*}
		\pcoll(\ell,p)^t&\leq \pcoll(1,p)^t\\
		&=(1-2p(1-p))^t\\
		&\leq (1-p)^t\\
		&\leq e^{-pt}\\
		&=e^{-w/2},
	\end{align*}
	where the first inequality follows from Lemma~\ref{lem:renyidec}, and the second inequality holds for large $n$ because $p=o(1)$.
	Therefore, for $\ell\leq \frac{w}{\log^2 n}$ it holds that
	\begin{align*}
		\binom{k}{\ell} \binom{n-k}{\ell} \pcoll(\ell,p)^t&\leq n^{\frac{2w}{\log^2 n}}\cdot e^{-w/2} =o(1/k).
	\end{align*}
	It is now enough to show that
	\begin{equation*}
		\binom{k}{\ell} \binom{n-k}{\ell} \pcoll(\ell,p)^t=o(1/k)
	\end{equation*}
	for all $\ell>\frac{w}{\log^2 n}$.
	Using Lemma~\ref{lem:ubcoll} along with standard upper bounds on binomial coefficients, for $n$ large enough we have
	\begin{align*}
		\binom{k}{\ell} \binom{n-k}{\ell} \pcoll(\ell,p)^t&\leq e^{2\ell}\left(\frac{k(n-k)}{\ell^2}\right)^\ell \left(\frac{1}{2\ell p(1-p)}\right)^{t/2}\\
		&\leq e^{2\ell+2\ell\log(n/\ell)}\cdot e^{-\frac{t\log(\ell p)}{2}}\\
		&\leq e^{2k(1+\log(n/k))-\frac{t\log(\ell p)}{2}}\\
		&=o(1/k).
	\end{align*}
	The third inequality follows from the fact that $\ell\leq k\leq n/e$. 
	The last equality follows from the constraints on $k$ and $w$ and the choice of $t$, since then for large enough $n$ and $c_\delta=2+1/\delta+\gamma$ we have
	\begin{align*}
		\frac{t\log(\ell p)}{2}&=\frac{t}{2}\cdot \log\left(\frac{w^2}{2t\log^2 n}\right)\\
		&=t\log w-\frac{t\log t}{2}-o(k\log(n/k))\\
		&\geq c_\delta k(1+\log(n/k))\\
		&\qquad -\frac{c_\delta k(1+\log(n/k))\log k}{2\log w}-o(k\log(n/k))\\
		&\geq c_\delta\left(1-\frac{1}{1+2\delta}\right)k(1+\log(n/k))-o(k\log(n/k))\\
		&\geq(2+\gamma')k(1+\log(n/k))
	\end{align*}
	for some constant $\gamma'>0$.
	The second inequality follows from the fact that $\frac{\log k}{\log w}\leq \frac{2}{1+2\delta}$.
	The last inequality follows by the definition of $c_\delta$.
	This concludes the proof.
\end{proof}

As mentioned before, Theorem~\ref{thm:wconsQNAGT} shows the lower bounds derived in Section~\ref{sec:QNAGTlb} are tight for $w$-constrained QNAGT schemes for a sizeable regime of $w$ and $k$.
This complements results for the unconstrained case presented in~\cite{SC17Phase,GHKL19}.
It remains an interesting open problem to verify whether our lower bounds are also tight in the runlength-constrained scenario.

\section*{Acknowledgment} The work was supported by the NSF Grant 1618366, the SemiSynBio NSF+SRC program under grant number 1807526 and the DARPA Molecular Informatics program.

\bibliographystyle{IEEEtran}
\bibliography{gt-refs-OM.bib}

\end{document}